\documentclass[reprint, superscriptaddress, amsmath,amssymb, aps, prresearch, longbibliography, floatfix]{revtex4-2}
\usepackage{amsmath}
\usepackage{graphicx}
\usepackage[normalem]{ulem}
\usepackage{bm}
\usepackage{natbib}
\usepackage{dsfont}
\usepackage{tikz}
\usepackage{epsfig}
\usepackage{feynmf}
\usepackage{blindtext, rotating}
\usepackage{mathtools}
\usepackage{dsfont}
\usepackage{subcaption}
\usepackage{physics}
\usepackage{amsfonts}
\usepackage{xcolor}
\usepackage{ragged2e}
\usepackage{siunitx}
\usepackage{comment}
\usepackage{soul}
\usepackage{empheq}
\usepackage{lipsum}
\usepackage{hyperref} 
\hypersetup{breaklinks=true, colorlinks=true, citecolor=blue, linkcolor=cyan, urlcolor=blue,filecolor=blue}
\usepackage{orcidlink}

\usepackage[T1]{fontenc}

\newcommand{\beginsupplement}{%
    \setcounter{table}{0}
    \renewcommand{\thetable}{S\arabic{table}}%
    \setcounter{figure}{0}
    \renewcommand{\thefigure}{S\arabic{figure}}%
    \setcounter{section}{0}
    \renewcommand{\thesection}{S\arabic{section}}%
    \setcounter{equation}{0}
    \renewcommand{\theequation}{S\arabic{equation}}%
}

\usepackage{xcolor} 

\DeclareCaptionJustification{justified}{\justifying}

\DeclareSIUnit{\rad}{rad}

\definecolor{bright_blue}{HTML}{85C1E9}
\definecolor{middle_blue}{HTML}{2E86C1}
\definecolor{dark_blue}{HTML}{1B4F72}

\begin{document}

\title{All-optical saddle trap as a platform for mesoscopic quantum experiments}


\author{Pedro V. Paraguassú \orcidlink{0000-0003-2334-5688}}%
\email{venturaskate@gmail.com}
\affiliation{Department of Physics, Pontifical Catholic University of Rio de Janeiro, Rio de Janeiro 22451-900, Brazil}

\author{Luca Abrahão \orcidlink{0009-0005-6266-5192}}%
\affiliation{Department of Physics, Pontifical Catholic University of Rio de Janeiro, Rio de Janeiro 22451-900, Brazil}

\author{Daniel Tandeitnik \orcidlink{0000-0003-3276-9335}}
\affiliation{Department of Physics, Pontifical Catholic University of Rio de Janeiro, Rio de Janeiro 22451-900, Brazil}

\author{D. Martínez-Tibaduiza \orcidlink{0000-0002-3257-6448}} 
\affiliation{Instituto de Física, Universidade Federal Fluminense, Niterói, Rio de Janeiro 24210-346, Brazil}

\author{Antonio Zelaquett Khoury \orcidlink{0000-0002-7487-5067}}
\affiliation{Instituto de Física, Universidade Federal Fluminense, Niterói, Rio de Janeiro 24210-346, Brazil}

\author{Thiago Guerreiro \orcidlink{0000-0001-5055-8481}}%
\email{thguerreiro@gmail.com}
\affiliation{Department of Physics, Pontifical Catholic University of Rio de Janeiro, Rio de Janeiro 22451-900, Brazil}

\begin{abstract}
We investigate the quantum dynamics of a levitated nanoparticle in a structured light rotating saddle-like optical potential consisting of a superposition of Gaussian and Laguerre-Gauss modes with detuned frequencies. This rotating saddle trap offers unique opportunities for quantum experiments, such as reduced decoherence due to photon recoil and absorption, the possibility of large delocalization of the particle's center-of-mass motion, particle recovery protocols, the generation of motional entanglement and momentum squeezing. As an application, we show that this saddle-trap architecture enables force detection with sensitivity in the zepto-Newton regime.


\end{abstract}


\maketitle

\textit{Introduction.}--- Levitated mechanical systems offer a unique platform for quantum science. The levitation of mesoscopic particles, in both optical and electric traps, has seen steady progress towards the quantum regime, with demonstrations of ground-state cooling \cite{delic2020cooling, magrini2021real, tebbenjohanns2021quantum, ranfagni2022two, kamba2022optical, piotrowski2023simultaneous, dania2025high}, center-of-mass (CoM) delocalization \cite{muffato2025generation, bonvin2024state} and quantum squeezing \cite{rossi2025quantum, mattana2026trap}, inter-particle interactions \cite{rieser2022tunable, vijayan2024cavity, deplano2024coulomb}, force metrology \cite{skrabulis2026nanomechanical, marocco2026three}, and fundamental tests of quantum mechanics in the mesoscopic realm \cite{pedalino2026probing}. Together, the field holds great promise for both fundamental \cite{aggarwal2022searching, monteiro2020search, aspelmeyer2022zeh, neumeier2024fast, bose2509spin, bose2025massive} and applied science \cite{rademacher2020quantum, millen2020optomechanics, gieseler2021optical, homans2025experimental}, but a number of important challenges must be overcome to promote levitated mesoscopic objects from proof-of-principle experiments to a mature quantum technology.

Traditionally, levitated optomechanics relies on Gaussian optical tweezers or Paul traps \cite{ashkin1976optical, paul1990electromagnetic}. While optical traps provide high-quality mechanical modes and an exquisite degree of isolation, they are affected by photon recoil heating \cite{jain2016direct} and the emission of blackbody radiation due to elevated bulk temperatures \cite{hackermuller2004decoherence, hebestreit2018measuring}. Together, these form the main sources of decoherence for quantum experiments with optically levitated nanoparticles in high vacuum. Paul traps, on the other hand, have lower frequencies and larger trap volumes \cite{bonvin2024hybrid, dania2024ultrahigh}, and the preparation of high-purity motional states of mesoscopic objects in these traps remains to be demonstrated \cite{dania2021optical}. Moreover, unlike elementary particles, mesoscopic objects are highly distinguishable; therefore, the ability of retrieving the same particle when performing a quantum experiment is a highly desirable feature, in order to avoid the washing-out of interference and other quantum effects \cite{clauser1997broglie}.


In this work, we investigate whether we can overcome these challenges and improve optomechanical quantum experiments by structuring the light field used to levitate particles while retaining the attractive features of optical levitation. For early explorations on the use of structured light in levitated optomechanics and optical traps see \cite{dorn2003sharper, dinter2024three, gomez2025optical, almeida2025levitated, almeida2023trapping, melo2020optical}. Here we explore the quantum dynamics of a mesoscopic particle within a novel type of structured light tweezer consisting of a saddle-like optical potential created by the superposition of three Laguerre-Gauss (LG) modes, referred to as the saddle beam \cite{tandeitnik2024all}. By changing the relative phase of these LG components -- for example, through frequency detunings -- the saddle can be made to rotate with a tunable angular velocity reaching the MHz range. This configuration enables the stable trapping of nanoscopic particles in a dynamical equilibrium, functioning as an all-optical analogue to a Paul trap, but with reduced trap volume and a unique set of characteristics, which we now explore in the context of levitated quantum experiments. 

\textit{Saddle trap dynamics.}--- We consider a family of optical trapping potentials generated in the paraxial approximation by a coherent superposition of three LG beams, whose total electric field $E_{tw}$ reads
\begin{equation}
    \frac{E_{tw}}{E_0} 
    = \sqrt{I_G}\, E_{0,0}^{\text{LG}}
      + \sqrt{\frac{I_L}{2}}\, E_{0,2}^{\text{LG}} e^{-i2\theta}
      + \sqrt{\frac{I_L}{2}}\, E_{0,-2}^{\text{LG}} e^{i2\theta},
    \label{trapping-field}
\end{equation}
where $E_{p,l}^{\text{LG}}$ denotes the Laguerre–Gaussian mode with radial index $p$ and azimuthal index $l$, and $E_0 = \sqrt{2P/(c\,\epsilon_0)}$ is the global normalization factor associated with the total optical power $P$. The parameter $\theta$ represents a tunable relative phase between the constituent LG modes. The total optical power in the trap is partitioned among the LG components such that $I_G$ and $I_L$ denote the fractional powers carried by the Gaussian ($l = 0$) and the Laguerre ($l = \pm 2$) modes, respectively, subject to the constraint $I_G + I_L = 1$. Consequently, the structure of the resulting optical potential can be controlled by varying the parameters $I_G$, $I_L$, and $\theta$.


When the phase parameter is varied at a constant rate, $\theta = \Omega t$, for instance by superposing the LG modes with slightly detuned frequencies \cite{tandeitnik2024all}, the field in Eq.~\eqref{trapping-field} generates a rotating saddle-shaped potential in the transverse $xy$-plane in the vicinity of the optical axis. Near the origin, the longitudinal motion is well approximated as harmonic and is effectively decoupled from the transverse dynamics, thereby enabling three-dimensional confinement of a dielectric particle in the dipole regime, provided that the rotation rate $\Omega$ exceeds a critical threshold value $\Omega_0$ \cite{tandeitnik2024all}. Since the longitudinal and transverse degrees of freedom are decoupled for sufficiently small particle displacements, we henceforth restrict our analysis to the motion in the $xy$-plane, which is governed by the time-dependent potential
\begin{equation}
    V(x,y,t) = V_0 \left[k_-(t) x^2 + k_+(t) y^2 - k_{xy}(t)xy \right],\label{eq:trap potential}
\end{equation}
where $V_0 = \Re{\alpha} P/ (c \pi w_0^2 \epsilon_0)$, with $\alpha$ denoting the particle polarizability, $w_0$ the waist parameter of the $E^{LG}_{p,l}$ modes, and
\begin{eqnarray}
    k_{\pm}(t) &=& I_G \pm 2\sqrt{I_L I_G} \cos(2 \Omega t), \\
    k_{xy}(t) &=& 4\sqrt{I_L I_G} \sin(2 \Omega t).
\end{eqnarray}

The quadratic approximation leading to Eq.~\eqref{eq:trap potential} remains valid provided that the transverse particle displacements satisfy $d_{\perp}\ll w_{0}/\sqrt{1+2\sqrt{I_L/I_g}}$. Within this regime, the potential is well approximated by a quadratic form and nonlinear contributions to the dynamics can be neglected \cite{roda2024numerical}. A detailed characterization of the saddle potential is provided in Ref.~\cite{tandeitnik2024all}, while an analysis of the quadratic confinement and validity of the quadratic approximation is presented in the Supplemental Material~\ref{app:saddle}.



Throughout this work, we adopt the following protocol and initial conditions. Prior to the initial time $t = 0$, a nanoparticle of radius $R = \SI{50}{nm}$ is confined in a standard Gaussian optical tweezer characterized by transverse trapping frequencies of $\Omega_q\approx \SI{150}{kHz}$, with corresponding zero-point motional amplitudes $q_{\text{zpf}} = \sqrt{\hbar/(2m\Omega_q)}$ for $q \in \{x, y\}$. The particle is prepared in a low-occupation thermal state of the CoM motion with $\bar{n}_x = \bar{n}_y = 0.8$. At $t = 0$, this state is transferred into the saddle optical potential (saddle beam) and subsequently evolves according to the master equation
\begin{equation}
    \frac{d\hat \rho}{dt} = -\frac{i}{\hbar} [\hat H, \hat \rho] 
    - \frac{\Gamma}{x_{\text{zpf}}^{2}} [\hat x, [\hat x, \hat \rho]]
    - \frac{\Gamma}{y_{\text{zpf}}^{2}} [\hat y, [\hat y, \hat \rho]],
    \label{eq: masterequation}
\end{equation}
where $\hat H$ denotes the Hamiltonian of the particle, incorporating the trapping potential defined in Eq.~\eqref{eq:trap potential}. The double-commutator terms represent Markovian decoherence processes, with $\Gamma$ as the decoherence rate, and $x_{\text{zpf}}$ and $y_{\text{zpf}}$ denoting the particle's zero-point motions corresponding to the Gaussian tweezer, that induce spatial localization of the particle along the $x$ and $y$ directions \cite{weiss2021large}. 

\begin{figure}[ht!]
    \centering
    \includegraphics[width=1\linewidth]{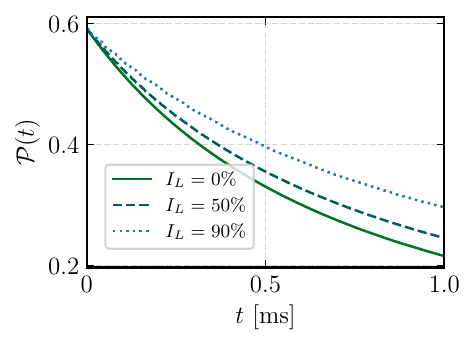}
    \caption{Time evolution of the state purity, $\mathcal{P}$ for different fractional powers $I_{L}$ of the $l=\pm 2$ modes in the saddle trap, for an initial thermal state of a Gaussian trap with occupation numbers $\bar{n}_{x} = \bar{n}_{y} = 0.8$ and initial purity $\mathcal{P}(0) \approx 0.6$. }
    \label{fig2purity}
\end{figure}

The main mechanisms of particle localization -- i.e. the main sources of decoherence -- are collisions with residual gas molecules, photon recoil heating \cite{jain2016direct}, and the emission of thermal radiation associated with elevated bulk temperatures \cite{hackermuller2004decoherence}. In ultra-high-vacuum conditions (pressures below $10^{-9}$\,mbar), the collisional decoherence rate is approximately $\Gamma_{\text{gas}} \approx \SI{100}{Hz}$, corresponding to a characteristic coherence time of $T_{\text{gas}} \approx \SI{10}{ms}$. This coherence time sets an upper bound on the duration of quantum experiments for which collisional decoherence can be safely neglected. By contrast, photon recoil heating and decoherence induced by thermal emission arise from the scattering and absorption of photons from the trapping beam. These mechanisms are intrinsically linked to the optical confinement and are therefore not readily mitigated in conventional optical traps. However, as we demonstrate below, both contributions can be substantially suppressed when employing the saddle trap.


\textit{Photon recoil.}--- By appropriately tuning the fractional powers of the Gaussian and LG modes, the photon recoil experienced by a trapped particle within the saddle-beam configuration can be substantially reduced. Heuristically, this behavior arises because the $l = \pm 2$ components of the trapping field \eqref{trapping-field} exhibit an intensity minimum (dark region) that does not contribute to photon recoil at leading order in the particle displacement; see Supplemental Material~\ref{app:recoilDec} for further discussions. The photon recoil decoherence rate for direction $q$ stemming from the Gaussian component of the saddle trap reads \cite{jain2016direct},
\begin{eqnarray}
    \Gamma_q = C_q \frac{I_G Pk^4\vert\alpha\vert^2}{6m\Omega_q^2w_0^2\pi^2\epsilon_0},
\end{eqnarray}
\noindent where
\begin{equation}\label{eq:geometriFactor}
    \mathbf{C}=\left[\frac{1}{5}, \frac{2}{5}, A^2+\frac{2}{5}\right]
\end{equation}
\noindent is a geometric factor due to the anisotropic way energy is transferred to the particle via photon recoil~\cite{PhysRevA.102.033505}, with $A$ depending on the trapping lens numerical aperture~\cite{tebbenjohanns2019optimal}.

\begin{figure}[ht!]
    \centering
    \includegraphics[width=1.0\linewidth]{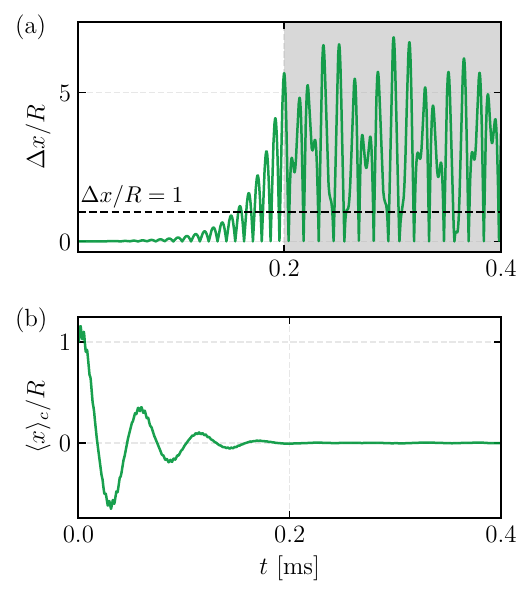}
    \caption{Time evolution of (a) the position uncertainty normalized by the particle radius, $\Delta x/R$. At $t = 0$, the particle is transferred into a slowly rotating saddle beam ($\Omega/\Omega_0 < 1$), leading to state expansion. At $t = \SI{0.2}{ms}$, the rotation frequency is abruptly increased to $\Omega/\Omega_0 = 4.5$, truncating state expansion. (b) Time evolution of the conditional mean position $\langle x \rangle/R$ under continuous monitoring and feedback cooling, starting from an initial displacement $\langle x \rangle_0 = R$.}
    \label{fig:statekalman}
\end{figure}

To quantify the suppression of recoil-induced heating in the saddle beam, we evaluate the time-dependent purity of the motional quantum state of the levitated particle, defined as $\mathcal{P}(t) = \mathrm{Tr}(\hat{\rho}^2)$. Figure~\ref{fig2purity} displays the temporal evolution of the purity for a saddle beam with wavelength $\lambda = \SI{1.55}{\mu m}$, total optical power $P = \SI{70}{mW}$, numerical aperture $\mathrm{NA} = 0.6$, and rotation frequency $\Omega / \Omega_0 = 4.5$, for various values of the fractional power $I_{L}$. As $I_{L}$ approaches $100\%$, the decay of purity becomes markedly slower, remaining bounded by $\mathcal{P}(t) \gtrsim 0.3$ over a characteristic gas-limited timescale $T_{\text{gas}} \approx \SI{1}{ms}$ for $I_L = 90\%$. This behavior contrasts sharply with that of conventional Gaussian optical traps, in which the initial state rapidly decoheres into low purity state over the same timescale. The diminished Gaussian contribution to the saddle-beam intensity profile, and the consequent reduction of photon-recoil events, represents a distinctive advantage of tailoring the spatial structure of the trapping field: although the total photon flux participating in the trap is the same as that of a standard optical tweezer, substantially less positional information about the particle is encoded in the scattered light and leaks out to the environment. Such suppression comes from the Gaussian mode being only a fractional component ($I_G$) of the total field and its lower numerical aperture ($\mathrm{NA} = 0.6$) compared to standard tweezers ($\mathrm{NA} \lesssim 1$). Since the recoil heating rate scales with $\mathrm{NA}^2$ \cite{jain2016direct}, a purely Gaussian trap at higher NAs would undergo a significantly more rapid purity decay.

\textit{Thermal emission}-- One of the main advantages of the proposed saddle beam configuration is the reduction of optical power absorbed by the particle. This arises from the fact that the majority of the beam intensity is distributed among Laguerre–Gaussian modes possessing a dark central intensity. Consequently, in comparison with a conventional Gaussian optical trap, the absorbed power is diminished and the bulk temperature of the particle is lowered. This reduction in temperature leads to a decreased rate of thermal radiation, thereby suppressing a potentially significant decoherence channel. The ratio of the absorbed power for the Saddle and Gaussian trapping configurations can be obtained by integrating the corresponding field intensities over the effective cross-sectional area of the particle, denoted \(A_p\). We have,
\begin{equation}
    \eta_{\mathrm{abs}} 
    = \frac{\displaystyle\int_{A_p}\bigl\lvert E_{tw}(\mathbf{r})\bigr\rvert^2\,\mathrm{d}^2\mathbf{r}}
           {\displaystyle\int_{A_p}\bigl\lvert E_{0,0}^{\text{LG}}(\mathbf{r})\bigr\rvert^2\,\mathrm{d}^2\mathbf{r}} \approx 1-I_L.
\end{equation}
\noindent where the approximation is valid in the dipole limit \(R~\ll~\lambda\).

\textit{State expansion and recovery}-- Reduced decoherence permits the implementation of quantum experiments over substantially extended timescales. An important class of such experiments that can be done using the saddle beam geometry concerns quantum particle delocalization, also referred to as large-scale wavefunction expansion or macroscopic state expansion.

Building on the previous analysis, we assume that, prior to the initial time, the particle is prepared in a motional state with a low phonon occupation number in a conventional Gaussian optical tweezer, under the same conditions as those previously employed to suppress recoil-induced decoherence. At $t = 0$, the particle is transferred into a slowly rotating saddle beam with rotation frequency $\Omega/\Omega_0 < 1$. Figure~\ref{fig:statekalman}(a) shows the temporal evolution of the variance of the particle’s $x$-coordinate, from which one infers that within $\SI{0.2}{ms}$ the particle undergoes pronounced delocalization, extending over several times its radius. 

To demonstrate that this delocalization is realized in a controlled manner, at $t = \SI{0.2}{ms}$ the saddle rotation frequency is abruptly increased to $\Omega/\Omega_0 = 4.5$, after which the variance remains bounded. This behavior confirms the capability of the system to release and subsequently re-capture the particle without the use of active feedback, relying solely on a modulation of the rotation frequency. Thereafter, the center-of-mass motion of the particle can be cooled via feedback cooling~\cite{magrini2021real,tandeitnik2024all}, as illustrated in Figure~\ref{fig:statekalman}(b), thereby returning the particle to a state close to its initial motional state and enabling repetition of the experimental sequence.

This protocol fulfills a central prerequisite for quantum experiments with mesoscopic objects, namely, the deterministic recapture of a delocalized object and the implementation of repeated measurements on the same individual particle.

\begin{figure}
    \centering
    \includegraphics[width=1\linewidth]{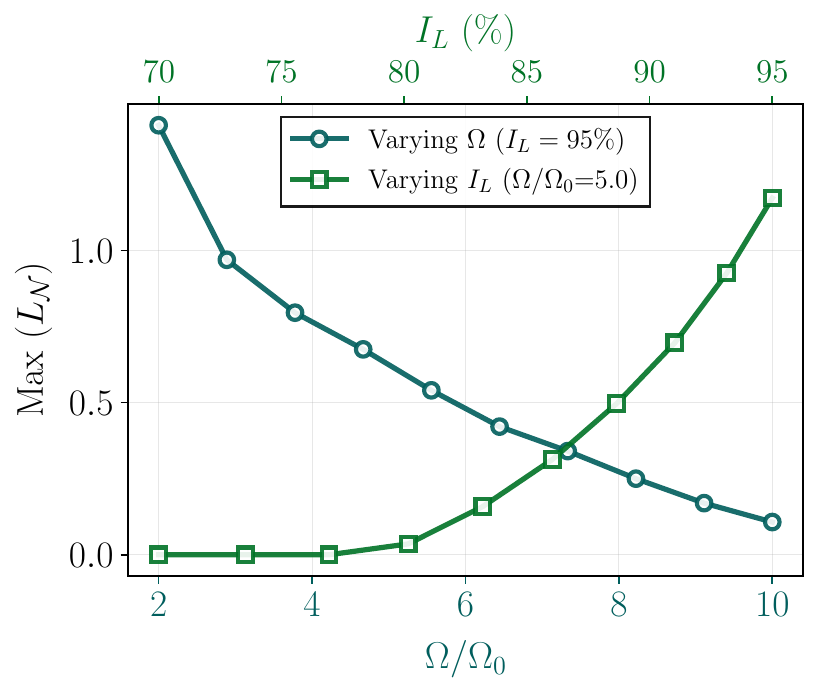}
    \caption{Maximum entanglement at 1~ms, quantified by the logarithmic negativity $L_{\mathcal{N}}$, as a function of the Laguerre-Gaussian fractional power $I_L$ and the dimensionless rotation frequency $\Omega/\Omega_0$. The lower and upper horizontal axes display the degree of rotation and the Laguerre mode fraction $I_L$, respectively.}
    \label{fig: entanglement}
\end{figure}

\textit{Entanglement.}--- We next discuss the generation of entanglement in the transverse $xy$ motional state of the levitated particle.   
This provides a clear demonstration of the potential of the saddle beam for generating non-classical states. 

When non-rotating, the saddle potential is highly non-isotropic, as can be seen by the presence of the cross-term in \eqref{eq:trap potential} proportional to $k_{xy}$, which is dependent on the fraction of $\ell = \pm2$ modes in the beam as $ k_{xy} \propto \sqrt{I_{G{}}} $. This term leads to entanglement between the $ x $ and $ y $ degrees of freedom of the particle's motion. We thus expect that motional entanglement between the $xy$ coordinates increases with $I_{G}$. This behavior can be seen in Fig. \ref{fig: entanglement} (squared dots), where we can see that the maximum logarithmic negativity of the particle's transverse motional state increases as the fraction of LG modes is increased at a fixed rotation frequency of $ \Omega = 5 \Omega_{0}$. Interestingly, we also observe that if we  maintain $I_{G}$ fixed, and increase the saddle rotation frequency, the maximum logarithmic negativity steadily decreases. This can be understood as an averaging effect: as the angular velocity increases, the trapped particle feels an ``average'' potential which becomes isotropic in the limit of large $\Omega$.

\begin{figure}
    \centering
    \includegraphics[width=1\linewidth]{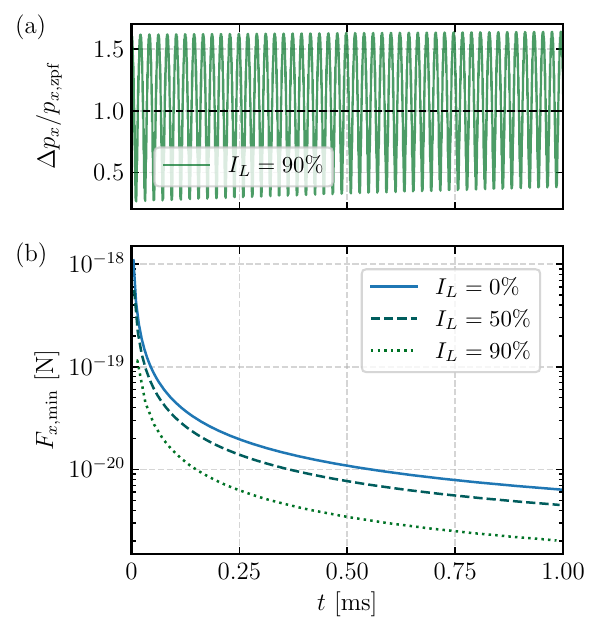}
    \caption{Time evolution of (a) the momentum uncertainty normalized by zero-point fluctuations of the initial Gaussian trap as discussed in the main text, $\Delta p_x / p_{\text{zpf}}$, specifically for $I_L = 90\%$. The dashed line marks the zero-point fluctuation limit, where values below it indicate momentum squeezing. (b) Minimum detectable force $F_{x, \text{min}}$ as a function of time for different Laguerre mode fractions $I_L$. The sensitivity is derived from the quantum Cramér-Rao Bound with a stable rotation frequency $\Omega = 20\,\Omega_0$. Increasing $I_L$ enhances momentum squeezing, directly improving sensitivity and pushing the detection limit into the zepto-Newton ($10^{-21}$~N) regime. }
    \label{fig:fisher}
\end{figure}

\textit{Force metrology.}--- Lastly, we evaluate the metrological capabilities of the saddle trap. As shown in Fig.~\ref{fig:fisher}(a), the system exhibits significant momentum squeezing, a key resource for enhancing force sensitivity. By increasing the Laguerre mode fraction $I_L$, we suppress decoherence, thereby preserving the squeezing necessary for precision measurements. To quantify this, we calculate the Quantum Fisher Information Matrix (QFIM) for constant forces acting along the transverse directions \cite{tsang2011fundamental, latune2013quantum}. Due to the system's symmetry, the sensitivity limits for $F_y$ are identical to those for $F_x$; thus, we focus our discussion on the latter. Using the Quantum Cramér-Rao bound with the Gaussian moments (see Supplemental Material), we derive the minimum detectable force shown in Fig.~\ref{fig:fisher}(b). The results demonstrate that higher $I_L$ values lead to superior sensitivity; notably, for $I_L=90\%$ , we achieve a detection limit in the tens of zepto-Newton regime ($10^{-20}\,\mathrm{N}$), highlighting the setup's potential for ultrasensitive force metrology. 

\textit{Conclusion.}--- In conclusion, the saddle beam provides several attractive features for levitated quantum optomechanical experiments. These include fast potential modulation through space-time structuring of light fields, reduced photon recoil, and lower bulk temperatures and decoherence due to absorption of light and re-emission of thermal photons. Furthermore, the system allows for large delocalizations comparable to the particle's size, particle recovery for large amplitudes of motion, tunable generation of motional entanglement and squeezing of the CoM degree of freedom, both in position and momentum. Furthermore, this motional entanglement and squeezing can find interesting applications in force metrology. In summary, through introducing structure in the optical field, the saddle beam offers a novel and complete toolbox for levitated quantum experiments.


\acknowledgments{
We acknowledge support from the Coordenac\~ao de Aperfei\c{c}oamento de Pessoal de N\'ivel Superior - Brasil (CAPES) - Finance Code 001, Conselho Nacional de Desenvolvimento Cient\'ifico e Tecnol\'ogico (CNPq), Funda\c{c}\~ao de Amparo \`a Pesquisa do Estado do Rio de Janeiro (FAPERJ Scholarship No. E-26/200.251/2023, E-26/210.249/2024 and FAPERJ PhD merit fellowship - FAPERJ Nota 10, 203.709/2025), Funda\c{c}\~ao de Amparo \`a Pesquisa do Estado de São Paulo (FAPESP processo 2021/06736-5), the Serrapilheira Institute (grant No. Serra – 2211-42299) and StoneLab.}

\bibliography{main}

\pagebreak 

\onecolumngrid
\beginsupplement

\vspace{1em}

\begin{center}
    \large\bfseries Supplemental Material
\end{center}
\vspace{1em}


\section{quadratic Confinement validity}\label{app:saddle}

Throughout this work, we consider the same definitions of $E_{p,l}^{\text{LG}}$ modes as introduced in ~\cite{tandeitnik2024all}, and a variation of the saddle beam introduced in the same reference. The main difference between the electric field in Eq. \eqref{trapping-field} and the saddle trap in \cite{tandeitnik2024all} is the introduction of the weight coefficients $I_L$ and $I_G = 1 - I_L$, which modify the shape of the potential.
To achieve confinement and stability in the trap, the transverse intensity profile must rotate with constant angular velocity $\Omega$ ($\theta = \Omega t$) exceeding a critical value. We now investigate under which conditions -- i.e. for which values of $\Omega$ and $ I_{L}$ -- trapping of a dielectric dipole can be achieved using the electric field \eqref{trapping-field}.

As shown in Ref.~\cite{tandeitnik2024all}, the rotation frequency stability criterion is given by
\begin{equation}
    \vert\Omega_0\vert > \sqrt{\frac{-\gamma^2+\omega_y^2-\omega_x^2+\sqrt{(\gamma^2+\omega_x^2-\omega_y^2)^2+4\omega_x^2 \omega_y^2}}{2}}.
    \label{eq:stability_criteria}
\end{equation}

\begin{figure}[ht!]
    \centering
    \includegraphics[width=1\linewidth]{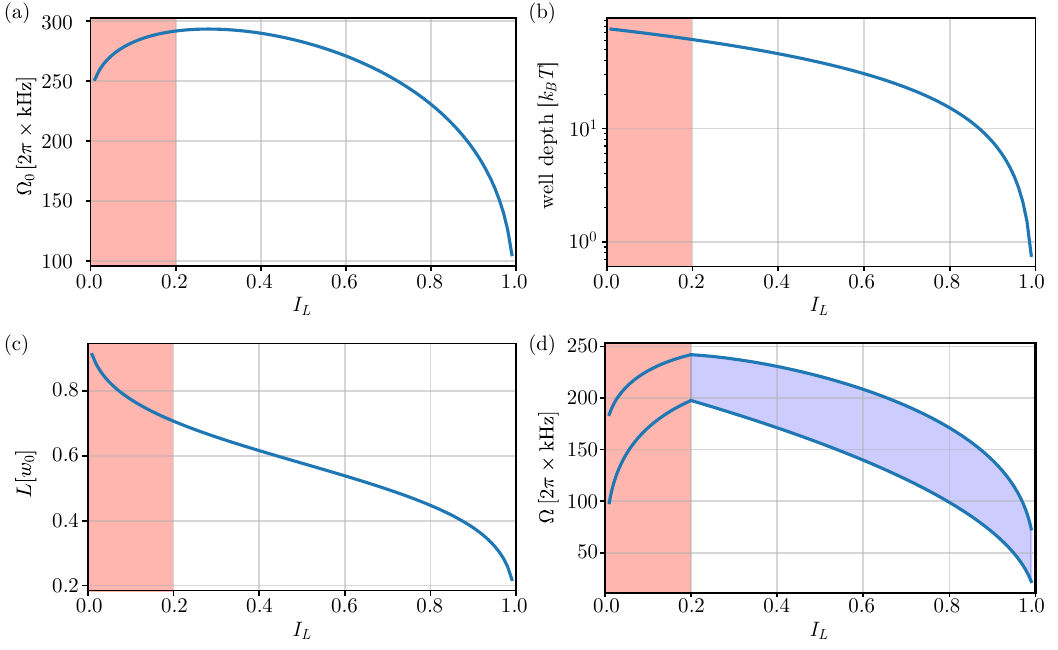}
    \caption{Dependence of the transverse confinement properties on the LG intensity fraction $I_L$. The red-shaded region marks $I_L\le 0.2$, where no saddle-like potential is formed. (a) Threshold rotation frequency $\Omega_0$ from Eq.~\eqref{eq:stability_criteria}. (b) Trap depth at the origin. (c) Characteristic length scale $L$ that determines the validity of the quadratic approximation. (d) Fast transverse oscillation beating frequency. Parameters correspond to $R=\SI{50}{nm}$.}
    \label{fig:ILpanelStudy}
\end{figure}

\noindent Moreover, the confinement properties of the potential given by \eqref{trapping-field} depend strongly on the LG intensity fraction $I_L$, as summarized in Fig.~\ref{fig:ILpanelStudy}. For $I_L \le 0.2$ (red-shaded region), no saddle-like potential exists. For $I_L > 0.2$, a saddle emerges. As $I_L$ increases further, the critical threshold frequency $\Omega_0$ decreases due to the flattening curvature along the unstable axis (Fig.~\ref{fig:ILpanelStudy}a). Concomitantly, the overall trap depth approaches zero as the origin transitions into a dark spot (Fig.~\ref{fig:ILpanelStudy}b).

It is also important to verify under which conditions the trapping potential can be approximated as a quadratic polynomial in th particle's transverse coordinates. As $I_L$ increases, the quartic contribution to the trapping potential becomes more prominent. The quadratic approximation of the trap remains valid only when the transverse motion amplitudes are strictly much smaller than the characteristic length scale $L$, defined by
\begin{equation}
    L = \frac{w_{0}}{\sqrt{1+2\sqrt{I_L/I_G}}}.
\end{equation}
As shown in Fig.~\ref{fig:ILpanelStudy}(c), even for highly elevated fractions such as $I_L = 90\%$ and larger, $L$ remains on the order of several hundred nanometers, thoroughly justifying the approximations used in the main text.

Finally, Fig.~\ref{fig:ILpanelStudy}(d) depicts the dependence of the principal transverse oscillation frequency on $I_L$. Here, we define the principal frequency as the arithmetic mean of the two neighboring low-frequency modes~\cite{tandeitnik2024all}, a choice that corresponds to the fast beating frequency which dominates the particle's dynamics. In the plot, we highlight the parameter range bounded from below by $\Omega/\Omega_0 = 1$ and extending to an upper regime characterized by $\Omega/\Omega_0 \gg 1$. As the ratio $\Omega/\Omega_0$ increases, the fast beating frequency converges to an approximately constant value for a given $I_L$, owing to the fact that the two neighboring frequencies move closer together and asymptotically approach an intermediate value between them~\cite{tandeitnik2024all}.

\section{Photon recoil decoherence}\label{app:recoilDec}

Photon recoil heating originates from the interaction between the trapping field and the vacuum (free) electromagnetic field, and gives rise to the recoil–heating contribution to the Hamiltonian
\begin{equation}
    H_{\text{rc}} = - \alpha \,\mathbf{E}_{\text{tw}}(\mathbf{r}) \cdot \mathbf{E}_{\text{f}}(\mathbf{r}),
\end{equation}
where the trapping field is defined in Eq.~\eqref{trapping-field}, and the free electromagnetic field is represented as
\begin{eqnarray}
    \mathbf{E}_{\text{f}}(\mathbf{r}) 
    = i \sum_{\epsilon} \int d^{3}\mathbf{k} \sqrt{\frac{\hbar \omega}{16\pi^{3}\varepsilon_{0}}}\, \mathbf{u}_{\epsilon}(\mathbf{k}) 
    \left[ a_{\epsilon}(\mathbf{k}) e^{i\mathbf{k}\cdot \mathbf{r}}  - a_{\epsilon}^{\dagger}(\mathbf{k}) e^{-i\mathbf{k}\cdot \mathbf{r}} \right].
\end{eqnarray}
Since we are interested in small displacements of the particle around its equilibrium position, $\boldsymbol{r} \approx 0$, we perform a Taylor expansion of both the free field and the trapping field about the origin and retain only the terms up to first order in the particle displacement.

From the explicit expression of, for example, the mode \(E^{LG}_{0,2}\),
\begin{equation}
    E^{LG}_{0,2}(r,\phi,z) = \sqrt{\frac{1}{\pi}}\frac{2r^2}{w^3(z)}\exp\left( -\frac{r^2}{w^2(z)} - i k\frac{r^2}{2R(z)} + i2\phi + i3\psi(z)\right),
\end{equation}
it is apparent that the lowest–order radial dependence already scales as \(r^2\), and thus does not contribute within the linear regime considered here. Consequently, only the Gaussian component of the saddle beam contributes to the recoil heating at first order in the particle displacement.

\section{Equations of motion for first and second moments}

In this section, we present the derivation of the first and second statistical moments for both the unconditional and conditional dynamical evolutions of the system.

\subsection{Unconditional Dynamics}
We describe the unconditional evolution of the system by the Lindblad master equation
\begin{equation}
    \frac{d\hat \rho}{dt} = -\frac{i}{\hbar} [\hat H, \hat \rho] 
    - \frac{\Gamma}{x_{\text{zpf}}^{2}} [\hat x, [\hat x, \hat \rho]]
    - \frac{\Gamma}{y_{\text{zpf}}^{2}} [\hat y, [\hat y, \hat \rho]].
\end{equation}
Because the potential $V(x,y)$ is quadratic in the position coordinates $x$ and $y$, the resulting dynamics is Gaussian. Consequently, the state of motion is fully characterized by the first moments and the covariance matrix of the canonical variables. We represent the state in phase space via the Wigner transform,
\begin{equation}
    W(x,y,p_x,p_y) = \frac{1}{(2\pi\hbar)^2} \iint_{-\infty}^{\infty} e^{-i (p_x s_x + p_y s_y) / \hbar} \,
    \big\langle x + \tfrac{s_x}{2}, y + \tfrac{s_y}{2} \big| \hat{\rho} \big| x - \tfrac{s_x}{2}, y - \tfrac{s_y}{2} \big\rangle \, ds_x \, ds_y,
\end{equation}
for which the time evolution is governed by a Fokker–Planck equation of the form
\begin{equation}
    \frac{\partial W}{\partial t} = \left( \mathcal{L}_{\text{Liouville}} W + \mathcal{L}_{\text{recoil}} W \right),
\end{equation}
where the Liouvillian and recoil superoperators, $\mathcal{L}_{\text{Liouville}}$ and $\mathcal{L}_{\text{recoil}}$, are given by
\begin{align}
    \mathcal{L}_{\text{Liouville}} W &= \{H, W\}_{\text{PB}} 
    = \left( \frac{\partial H}{\partial x}\frac{\partial W}{\partial p_x} + \frac{\partial H}{\partial y}\frac{\partial W}{\partial p_y} \right)
      - \left( \frac{\partial H}{\partial p_x}\frac{\partial W}{\partial x} + \frac{\partial H}{\partial p_y}\frac{\partial W}{\partial y} \right), \\
    \mathcal{L}_{\text{recoil}} W &= \hbar^2 \Gamma \left( \frac{1}{x_{\text{zpf}}^{2}} \frac{\partial^2 W}{\partial p_x^2} + \frac{1}{y_{\text{zpf}}^{2}} \frac{\partial^2 W}{\partial p_y^2} \right).
\end{align}
To obtain the equations of motion for the statistical moments, we evaluate the time derivative of the expectation value of the phase-space variables. For a general observable $O(x, y, p_x, p_y)$, this is expressed as an integral over phase space involving the time evolution of the Wigner function,
\begin{equation}
    \frac{d\langle O \rangle}{dt} = \int_{-\infty}^{\infty} O \frac{\partial W}{\partial t} \, dx \, dy \, dp_x \, dp_y.
\end{equation}
Upon substituting the Wigner equation into this expression and performing integration by parts—under the assumption that the Wigner function and its derivatives vanish at infinity—the differential operators can be transferred from $W$ to the observable $O$. In this representation, the Liouvillian term reproduces the classical Hamiltonian flow in phase space, whereas the recoil term manifests as an effective momentum-diffusion process, contributing exclusively to the momentum variances with rates given by $\Lambda_x = \Gamma/x_{\text{zpf}}^{2}$ and $\Lambda_y = \Gamma/y_{\text{zpf}}^{2}$.

This property enables a substantial reduction of complexity: rather than solving the full master equation for the density matrix, the system’s dynamics are completely characterized by the evolution of the first and second statistical moments. The equations of motion for the first moments (expectation values) read
\begin{align}
    \frac{d\langle x\rangle}{dt} &= \frac{1}{m} \langle p_x \rangle, & \frac{d\langle y\rangle}{dt} &= \frac{1}{m} \langle p_y \rangle, \\
    \frac{d\langle p_x\rangle}{dt} &= - \left\langle \frac{\partial V}{\partial x} \right\rangle, & \frac{d\langle p_y\rangle}{dt} &= - \left\langle \frac{\partial V}{\partial y} \right\rangle.
\end{align}

The temporal evolution of the ten independent components of the symmetric covariance matrix is governed by the following system of coupled ordinary differential equations:
\begin{align}
    \frac{d\sigma^2_x}{dt} &= \frac{2}{m}C_{xp_x}, & \frac{d\sigma^2_y}{dt} &= \frac{2}{m}C_{yp_y}, \\
    \frac{d\sigma_{p_x}^2}{dt} &= -2 C_{\partial_x V, p_x}  + 2\hbar^2\Lambda_x, & \frac{d\sigma_{p_y}^2}{dt} &= -2 C_{\partial_y V, p_y}  + 2\hbar^2\Lambda_y, \\
    \frac{dC_{xy}}{dt} &= \frac{1}{m}\left(C_{xp_y}+C_{yp_x} \right), & \frac{dC_{p_x p_y}}{dt} &= -\left( C_{\partial_x V, p_y} + C_{\partial_y V, p_x}\right), \\
    \frac{dC_{xp_x}}{dt} &= \frac{1}{m}\sigma_{p_x}^2 - C_{\partial_x V, x}, & \frac{dC_{yp_y}}{dt} &= \frac{1}{m}\sigma_{p_y}^2  - C_{\partial_y V, y}, \\
    \frac{dC_{xp_y}}{dt}  &= \frac{1}{m}C_{p_x p_y } - C_{\partial_y V, x}, & \frac{dC_{yp_x}}{dt} &= \frac{1}{m}C_{p_x p_y } - C_{\partial_x V, y}.
\end{align}

Taken together, these equations constitute a closed system of 14 coupled ordinary differential equations that can be integrated numerically to reconstruct the full dynamical evolution of the system. The covariance terms involving the potential are defined in the usual way; for instance,
\begin{equation}
    C_{\partial_x V, x} = \langle (\partial_x V) x \rangle - \langle \partial_x V \rangle \langle x \rangle,
\end{equation}
with analogous definitions for all remaining second-moment contributions.

\subsection{Conditional Dynamics under Continuous Monitoring}

We now consider the situation in which the light scattered by the nanoparticle is continuously monitored in order to extract information about its center-of-mass position. Operationally, this is implemented by measuring the phase of the scattered field. The resulting conditional dynamics of the system, conditioned on the measurement record, are described by a stochastic master equation (SME) for the Wigner function $W(x, y, p_x, p_y)$,
\begin{align}
    dW ={}& \left( \mathcal{L}_{\text{Liouville}} W + \mathcal{L}_{\text{recoil}} W \right) dt \nonumber \\
    &+ \sqrt{2 \eta_x \Lambda_x} \left( x - \langle x \rangle \right) W \, d\zeta_x + \sqrt{2 \eta_y \Lambda_y} \left( y - \langle y \rangle \right) W \, d\zeta_y. \label{eq:sme_wigner}
\end{align}
Here, $\mathcal{L}_{\text{Liouville}}$ denotes the Liouvillian describing the coherent evolution under the trapping potential, while $\mathcal{L}_{\text{recoil}}$ accounts for momentum diffusion arising from recoil heating, both as defined previously. For notational simplicity, we assume symmetric measurement and decoherence rates along the two transverse axes, $\Lambda_x = \Lambda_y = \Lambda$ and $\eta_x = \eta_y = \eta$, where $\eta$ denotes the quantum efficiency of the detection process. 

The last two terms in Eq.~\eqref{eq:sme_wigner} encode the effect of a weak, continuous quantum measurement of the position coordinates. This measurement induces stochastic back-action, modeled by independent Wiener increments $d\zeta_i$ that satisfy the It\^o rules $\langle d\zeta_i \rangle = 0$ and $d\zeta_i d\zeta_j = \delta_{ij} \, dt$.

Employing It\^o calculus, we obtain the evolution equations for the statistical moments. The first moments (mean values) obey the following stochastic differential equations (SDEs):
\begin{align}
    d \langle x \rangle &= \frac{1}{m}\langle p_x\rangle \, dt + \sqrt{2\eta \Lambda} \left(\sigma_x^2 \, d\zeta_x + C_{xy}\, d\zeta_y\right), \\
    d \langle y \rangle &= \frac{1}{m}\langle p_y\rangle \, dt + \sqrt{2\eta \Lambda} \left(\sigma_y^2 \, d\zeta_y + C_{xy}\, d\zeta_x\right), \\
    d \langle p_x \rangle &= - \langle \partial_x V \rangle \, dt + \sqrt{2\eta \Lambda} \left(C_{xp_x} \, d\zeta_x + C_{yp_x} \, d\zeta_y\right), \\
    d \langle p_y \rangle &= - \langle \partial_y V \rangle \, dt + \sqrt{2\eta \Lambda} \left(C_{xp_y} \, d\zeta_x + C_{yp_y} \, d\zeta_y\right).
\end{align}
Importantly, the measurement back-action on the mean values is proportional to the elements of the covariance matrix. This dependence gives rise to a nonlinear feedback mechanism in which the instantaneous state uncertainty determines the effective strength of the measurement-induced noise acting on the conditional trajectory.

In contrast, the second moments evolve deterministically and satisfy a set of modified Riccati-type equations, which now include contributions from information gain due to the continuous measurement:
\begin{align}
    \frac{d\sigma_x^2}{dt} &= \frac{2}{m}C_{xp_x} - 2\eta \Lambda \left(\sigma_x^4+C_{xy}^2\right), \\
    \frac{d\sigma_y^2}{dt} &= \frac{2}{m}C_{yp_y} - 2\eta \Lambda \left(\sigma_y^4+C_{xy}^2\right), \\
    \frac{d\sigma_{p_x}^2}{dt} &= -2 C_{\partial_x V, p_x}  + 2\hbar^2\Lambda - 2\eta\Lambda \left(C_{xp_x}^2+C_{yp_x}^2\right), \\
    \frac{d\sigma_{p_y}^2}{dt} &= -2 C_{\partial_y V, p_y}  + 2\hbar^2\Lambda - 2\eta\Lambda \left(C_{xp_y}^2+C_{yp_y}^2\right), \\
    \frac{dC_{xy}}{dt} &= \frac{1}{m}\left(C_{xp_y}+C_{yp_x} \right) - 2\eta\Lambda C_{xy}\left(\sigma_x^2+\sigma_y^2\right), \\
    \frac{dC_{p_x p_y}}{dt} &= -\left( C_{\partial_x V, p_y} + C_{\partial_y V, p_x}\right) - 2\eta\Lambda\left(C_{xp_x}C_{xp_y}+C_{yp_y}C_{yp_x}\right), \\
    \frac{dC_{xp_x}}{dt} &= \frac{1}{m}\sigma_{p_x}^2 - C_{\partial_x V, x} - 2 \eta \Lambda \left( \sigma_x^2 C_{xp_x}+C_{xy}C_{yp_x}\right), \\
    \frac{dC_{yp_y}}{dt} &= \frac{1}{m}\sigma_{p_y}^2  - C_{\partial_y V, y}- 2 \eta \Lambda \left(\sigma_y^2 C_{yp_y}+C_{xy}C_{xp_y}\right), \\
    \frac{dC_{xp_y}}{dt} &= \frac{1}{m}C_{p_x p_y } - C_{\partial_y V, x} - 2\eta \Lambda\left(\sigma_x^2 C_{xp_y}+C_{xy}C_{yp_y}\right), \\
    \frac{dC_{yp_x}}{dt} &= \frac{1}{m}C_{p_x p_y } - C_{\partial_x V, y}- 2\eta \Lambda\left(\sigma_y^2 C_{yp_x}+C_{xy}C_{xp_x}\right).
\end{align}
Here, the terms proportional to $\eta\Lambda$ quantify the competition between measurement-induced localization (information gain) and the concomitant back-action on the conjugate variables, thereby determining the conditional covariance dynamics of the nanoparticle’s motional state.

\section{Dynamics under feedback cooling}

In this section, we analyze the conditional dynamics of the system in the presence of active feedback cooling. To counteract recoil heating and stabilize the particle’s motion, the continuous position measurement is exploited to implement a closed-loop feedback protocol. This is realized by applying an additional force, derived from a Kalman-filter-based estimate of the particle’s state, which actively damps its mechanical energy.

The feedback is incorporated into the dynamics via an additional linear term in the system Hamiltonian,
\begin{equation}
    H_{\text{total}} = H_{\text{trap}} + H_{\text{feedback}} = H_{\text{trap}} + u_x(t) x + u_y(t) y,
\end{equation}
where $H_{\text{trap}}$ denotes the unperturbed trapping Hamiltonian. The feedback control signals $u_x(t)$ and $u_y(t)$ are computed in real time from the optimal state estimate provided by the Kalman filter. Consequently, the stochastic differential equations for the first moments of the momenta acquire additional deterministic control contributions:
\begin{align}
    d \langle p_x \rangle &= - \langle \partial_x V \rangle \, dt - u_x(t) \, dt + \sqrt{2\eta \Lambda} \left(C_{xp_x} \, d\zeta_x + C_{yp_x} \, d\zeta_y\right), \\
    d \langle p_y \rangle &= - \langle \partial_y V \rangle \, dt - u_y(t) \, dt + \sqrt{2\eta \Lambda} \left(C_{xp_y} \, d\zeta_x + C_{yp_y} \, d\zeta_y\right),
\end{align}
while the equations governing the mean positions and the covariance matrix remain unchanged.

To quantify the performance of the feedback scheme, we consider a worst-case scenario in which the particle is initially prepared far from the trap center. Specifically, we initialize the system with a large mean position $\langle x \rangle(0) = 10^4 x_{\text{zpf}}$, while setting $\langle y \rangle(0)$ and all initial momenta to zero, and then activate the feedback protocol. 


Increasing $I_L$ suppresses recoil heating but simultaneously reduces the scattered-photon flux, thereby diminishing the information available to the Kalman filter. Despite this weaker measurement signal, the feedback loop robustly cools the particle back to its motional ground state in all examined cases. These results demonstrate that the feedback protocol functions as an effective “safety net,” capable of recovering the particle from large excursions in phase space and maintaining trap stability over a broad range of operating conditions.

\section{Quantum Fisher Information and Minimum Detectable Force}

In this section, we provide a detailed derivation of the Quantum Fisher Information (QFI) associated with force estimation and obtain the corresponding theoretical lower bound for the minimum detectable force acting on the nanoparticle.

We consider a nanoparticle subjected to weak, constant, and a priori unknown forces, $f_x$ and $f_y$, applied along the principal axes. The interaction is governed by the Hamiltonian
\begin{equation}
    H_{\text{int}} = -f_x x - f_y y.
\end{equation}
In the adopted dimensionless representation, this interaction takes the form
\begin{equation}
    H_{\text{int}} = -F_x \tilde{x} - F_y \tilde{y},
\end{equation}
where the dimensionless forces $F_x$ and $F_y$ are related to the corresponding physical forces (in Newtons) via
\begin{align}
    f_x &= F_x \frac{\hbar \Omega}{x_0}, & 
    f_y &= F_y \frac{\hbar \Omega}{y_0},
\end{align}
with $x_0$ and $y_0$ denoting the characteristic length scales along the $x$ and $y$ directions, respectively.

The ultimate precision for estimating these forces is constrained by the Quantum Cramér–Rao Bound (QCRB), which is expressed in terms of the Quantum Fisher Information matrix, $\mathcal{F}$. For a Gaussian state undergoing a displacement in phase space, the QFI matrix elements depend solely on the first and second statistical moments and are given by
\begin{equation}
    \mathcal{F}_{ij} = 2 \left( \frac{\partial \mathbf{d}}{\partial F_i} \right)^\top 
    \sigma^{-1} 
    \left( \frac{\partial \mathbf{d}}{\partial F_j} \right),
\end{equation}
where $\mathbf{d} = (\langle \tilde{x} \rangle, \langle \tilde{p}_x \rangle, \langle \tilde{y} \rangle, \langle \tilde{p}_y \rangle)^\top$ is the vector of first moments and $\sigma$ is the $4 \times 4$ symmetric covariance matrix of the quadrature operators.

The QCRB states that the variance of any unbiased estimator of the force parameter $F_i$ is lower-bounded by the corresponding diagonal element of the inverse QFI matrix:
\begin{equation}
    \mathrm{Var}(F_i) \geq (\mathcal{F}^{-1})_{ii}.
\end{equation}

A larger value of the QFI signifies a stronger sensitivity of the quantum state to changes in the force parameters, thereby enabling more precise parameter estimation. Consequently, the smallest physically resolvable force along the $x$-axis is given by the standard-deviation bound
\begin{equation}
    f_{x,\text{min}} = \sqrt{(\mathcal{F}^{-1})_{xx}} \, \frac{\hbar \Omega}{x_0}.
\end{equation}
An analogous expression holds for the $y$-axis. This theoretical framework demonstrates that any protocol capable of suppressing quantum noise in the relevant phase-space quadratures—such as momentum squeezing induced by structured optical fields—enhances the corresponding elements of the QFI matrix. As a result, the minimum detectable force is reduced and the fundamental metrological sensitivity of the system is improved.
\end{document}